# Measurement of the effects of rough surfaces on Lamb waves propagation

Damien Leduc[1], Bruno Morvan, Pascal Pareige, Jean-Louis Izbicki

*Laboratoire d'Acoustique Ultrasonore et d'Electronique (L.A.U.E.), UMR CNRS 6068,*

*Université du Havre, Place Robert Schuman, BP 4006, 76610 Le Havre, France*

**Abstract**

This work deals with the sensitivity to the plate roughness of Lamb waves. An experimental study is performed involving an air-coupling transducer system. Signal processing allows us to extract the Lamb waves characteristics: phase velocity and attenuation. Plate surface topographies are obtained by means of an optical surface profiler. The acoustic characteristics and the surface topographies are finally linked.

1. **Introduction**

   The use of ultrasonic waves to control adhesive bonding is widespread nowadays (see e.g. [1]). In most cases of an adhesive joint, the substrate surface is treated before application of the adhesive. The goal is to increase the interfacial energy in the contact

---

[1] Corresponding author: tel.: +332 32 74 47 18; fax: +332 32 74 47 19

  E-mail address: damien.leduc@univ-lehavre.fr

region. There are many different chemical or mechanical treatments; most of them consist in increasing the substrate surface roughness. Indeed, the surface roughness increases the area of physical contact and improves the adhesion strength. Estimation of the joint quality by means of acoustic methods cannot be performed without understanding the roughness effect on the ultrasonic waves propagation. In a bond structure, the plate rough side is located at the interface and there is of course no access to this surface for direct evaluation with surface waves (i.e. Rayleigh wave type). That is why Lamb waves seem accurate for future study. While these waves will be generated from the outer side of the plate, they may provide us with the bonding inner surface.

Many studies deal with roughness influence on ultrasonic waves reflection at a fluid-solid interface ([2], [3]). Few of them focus on guided waves propagation ([4], [5]). Brekhovskikh [6] has formulated an analysis that predicts the attenuation of Rayleigh waves travelling on a periodic rough surface over a particular frequency band. Chimenti and Lobkis [7] show that guided waves damping increases with surface roughness and frequency for an immersed plate. However roughness has no influence on the phase velocity of Lamb waves.

This work is an experimental study involving air-coupling measurement of Lamb waves on rough plates. Signal analysis allows us to extract, at a given frequency, the real ($k_x'$) and imaginary ($k_x''$) components of the wave vector of the Lamb waves propagating along the x direction. The real part $k_x'$ of the wave number gives the wave phase velocity whereas its imaginary part $k_x''$ gives the amplitude attenuation. We will study these parameters evolution versus the roughness parameter.

In a first part, the surface topography of many samples is studied. Then acoustic measurements are reported and discussed in the next section. Finally a brief conclusion is given.

## 2. Surface topography

### 2.1 Characterization of the studied samples

The studied samples are glass plates with *200 mm* sides and a thickness *d = 5 mm*. The measured acoustics parameters are the glass density: *2490 kg m$^{-3}$* and the shear and longitudinal velocities: $c_T$ = *3485 m s$^{-1}$* and $c_L$ = *5825 m s$^{-1}$*. These plates are treated on one side only in order to obtain a rough surface except one that is considered as a reference plate. The other surface is roughless and exactly the same for all samples: this surface stands for a reference surface. Two quite different techniques are used to create surface roughness: sanding and shot blasting. We may observe the studied surface topography by means of an optical surface profiler. The analysed surface is about *25 mm$^2$*. The resolution is *5 μm* in *x* and *y* directions and *10 nm* in *z* direction.

### 2.2 Quantitative surface evaluation by optical profiler

We characterize the roughness by use of the usual statistic parameters $R_a$, $R_q$ and $R_t$ [8].

The $R_a$ is called the roughness average; it is the arithmetic average of the absolute values of the measured height deviations. Height deviations $Z_{ij}$ are measured from the mean surface. $R_a$ mathematical expression is:

$$R_a = \frac{1}{MN} \sum_{j=1}^{M} \sum_{i=1}^{N} |Z_{ij}| \tag{1}$$

where *M* and *N* are the number of data points in each direction of the analysed surface.

$R_q$, also called the root mean square (*RMS*) roughness, is obtained by squaring each $Z_{ij}$ value over the evaluation area, then taking the square root of the mean:

$$R_q = \sqrt{\frac{1}{MN}\sum_{j=1}^{M}\sum_{i=1}^{N} Z_{ij}^2} \quad (2)$$

$R_t$ is the maximum peak-to-valley height; it is the vertical distance between the highest and lowest points of the surface. The statistic parameters of the four studied samples are reported in table 1. These mean values are calculated from six acquisitions on each plate providing from different zones. As shown in table 1, roughness values cover a large range of $R_q$.

3. **Acoustic measurements**

3.1 Experimental set-up

The experimental set-up is reported in figure 1. A pulse generator delivers a very short pulse voltage (about *300 V* during *300 ns*) to a Panametrics™ emitting piezo-composite transducer. Lamb waves are generated by the wedge method. The receiver transducer is a Sofratest™ air-coupling piezo-electric transducer. The air-coupling transducer detects the re-radiated signal. The emitter transducer remains unmoved whereas the receiver is translated along the propagation direction of Lamb waves. The use of a receiver air-coupling transducer is very interesting for such a study. Experiments are repeatable because we avoid all the problems due to the contact of the transducer with the sample (to maintain unchanged the thickness of coupling gel and the pressure on the transducer during the measurement). Moreover, air-coupling measurements enable motorized translation and automatic data acquisition.

The two transducers are set on the non-rough side. The displacement amplitudes are collected from *x = 10 mm* to *x = 90 mm* by *0.1 mm* step (origin *x = 0* corresponds to the wedge position). For each position of the air-coupling transducer, a *200* μs long signal is acquired on *10000* points. In order to improve the signal to noise ratio, we performed an average of *1000* successive shots.

3.2 Signal processing

Figure 2 is the result of a spatial followed by a temporal Fast Fourier Transform [9] of the temporal signals. Waves are generated by a pulse excitation so we obtain Lamb waves on a large frequency bandwidth. The receiving air-coupling transducer bandwidth is narrower than the emitting contact transducer one and allows signal detection for a product thickness-frequency belonging to the range *5* and *15 MHz mm*. Theoretical dispersion curves are superimposed on experimental data. Four Lamb modes may be identified: the A1 and A2 anti-symmetric modes and the S1 and S2 symmetric modes. The experimental wave numbers $k_x'$ for the reference sample (without roughness) are close to the theoretical ones. In the next paragraph, wave numbers of Lamb waves propagating in rough sample are compared to the reference ones.

3.2.1 Lamb waves $k_x'$ evolution for different surface roughness

The previous study for reference sample is now applied on the three plates with increasing roughness. We notice that, for the roughest sample, no mode is excited above $f\,d = 13\,MHz\,mm$. This is probably due to the attenuation of the high frequencies modes propagating in rough plates. At high frequencies, the wavelengths of the Lamb modes become commensurate with the size of the roughness. A multi-scattering process can occur and consequently the propagation of the Lamb modes cannot be detected. That is why the $k_x'$ comparison is performed only on the three first modes A1, S1 and A2. The experimental values of $k_x'$ are the maximum values of signals in *(k,f)* representation. In that way, we have evaluated $k_x'$ values of the Lamb waves for each studied sample and reported them in figure 3a. At a given frequency, the experimental $k_x'$ for each sample are close to one another. Figure 3b shows the results for the S1 mode: the shift between the theoretical and the experimental $k_x'$ increases with the roughness.

The relative difference defined by:

$$\frac{\Delta k_x'}{k_x'(ref)} = \frac{k_x'(rough) - k_x'(ref)}{k_x'(ref)} \qquad (3)$$

between rough and reference samples is reported in table 2. This difference does not significantly change from a mode to one another, in other words it is independent to frequency shifts. It increases however with roughness even if its variation is not linear. Indeed the $R_q$ coefficient for the shot blasting plate is *5* times higher than the sanding plate (table 1) whereas the relative difference is almost the same. On the contrary the $R_q$ value between the shot blasting and the strong shot blasting plates is twice greater and the relative difference increases from *2 %* to *4.6 %*. For the two smoothest samples, the relative difference variation is in the range of detectability. For the roughest plate, the value of the relative difference is outside the experimental error domain and it can be concluded that, for a large value of roughness, an effect on the value of $k_x'$ exists. Nevertheless the value of $k_x'$ does not look like a good criterion to discriminate different roughness. In the next paragraph, we will study the effect of roughness on Lamb waves attenuation.

### 3.2.2 Comparison of Lamb waves attenuation

In order to isolate one mode for its further characterisation, a spatial Short Time Fourier Transform (STFT) is achieved. First a temporal Fast Fourier Transform of the signal $s(x,t)$ is performed providing a $\tilde{s}_t(x,f)$ signal in the spatial-frequency domain. Then, the frequency corresponding to the studied mode is isolated using a spatial window sliding in the entire spatial domain (*80 mm*). A spatial FFT is computed for each position and provides a $\tilde{s}_{x,t}(k,f)$ signal. Thus the amplitude attenuation versus distance between the emitting and receiving transducer is visualized in the *(x,k)* space.

After the STFT treatment, the amplitude for each mode is obtained and plotted in figure 4 and 5 versus propagation distance. On the figures, the level one corresponds to initial position. For a given mode, this normalization allows us to compare the attenuation for different plates.

Firstly we focus on a particular mode and we study the evolution of the wave amplitude versus the propagation distance x for different roughness (figure 4). This evolution is almost linear when the roughness is weak ($R_q < 6\ \mu m$). Attenuation exists for the reference sample. This attenuation is close to the sanded sample one. It is mainly due to the diffraction phenomena occurring by the limited size of the emitter transducer. It has to be noticed that a small part of the energy is also re-radiated in the surrounding medium. This attenuation can be qualified as "apparent". With frequency increasing, the effect of the diffraction diminishes nevertheless this attenuation still exists for the rough samples. For a given frequency, the attenuation increases with increasing roughness (figure 4a, 4b and 4c). This fact has been experimentally verified for each of the considered mode.

Secondly the amplitude evolution of a selected roughness for different modes is shown in figure 5. We note that the higher frequency the higher the attenuation. By example, for a propagation distance of *70 mm*, the attenuation varies from *3.5 dB* (at *5.9 MHz mm*) to *6.9 dB* (at *13.1MHz mm*) for the shot blasted plate. The sensitivity to the roughness is then stronger in high frequency range.

As a partial conclusion, contrary to the $k_x^{'}$ value, the attenuation seems to be a good parameter in order to discriminate between different roughness.

## 4. Conclusion

After determining first the surface topography, secondly the effect of roughness of ultrasonic waves, we join the two parts of our work. Whereas the $k_x^{'}$ value is only slightly affected by the $R_q$ parameter as discussed in section 3.2.1, the attenuation is selected to be an efficient acoustic parameter. Thus we plot the attenuation versus $R_q$ statistic parameter for a given Lamb mode (figure 6). This representation links explicitly acoustic parameters and surface topography. Obviously a larger study involving more samples with a larger range of roughness would be necessary to obtain calibration curves.

Future works will concern adhesive interfaces in order to estimate the influence of the roughness on the bond quality.

**Acknowledgement**

The authors would like to thank Pascal Rembert for *Matlab ™* help, *Liot Techni-verres* for providing samples and *CRITT Analyses and Surfaces* for the surface topography.

# Table caption

**Table 1:** Roughness statistic parameters for the studied samples.

**Table 2:** $k_x^{'}$ relative difference values between rough and reference samples. For the strong shot blasted sample, the S2 mode is too attenuated.

# Figure caption

**Figure 1:** Experimental set-up.

**Figure 2:** Representation of the signal in the dual space *(k,f)* for reference sample. The lines correspond to the theoretical dispersion curves: solid lines for anti-symmetric modes and dotted lines for symmetric modes. Experimental data are in grey level. They practically coincide with the theoretical data.

**Figure 3:** Comparison of the $k'_x$ for different samples:
   a. For given mode, the $k'_x$ value are almost unchanged whatever the sample.
   b. Zoom on the S1 mode $k'_x$ values.

   A small effect on the $k'_x$ value is shown. Nevertheless this effect is the order of magnitude of the experimental error.

**Figure 4:** Evolution of Lamb waves attenuation versus propagation distance:
   a. for the A1 mode,
   b. for the S1 mode,
   c. for the A2 mode.

**Figure 5:** Evolution of Lamb waves attenuation for the shot blasted plate at several frequencies.

**Figure 6:** Evolution of the acoustic normalized amplitude versus $R_q$ surface parameter for different propagation distances (x = 30, 50 or 70 mm):
   a. A1 mode,
   b. S1 mode,
   c. A2 mode.

Tab. 1

|  | $R_a$ (µm) | $R_q$ (µm) | $R_t$ (µm) |
|---|---|---|---|
| **non treated** | < 0.01 | < 0.01 | 0.10 |
| **sanded** | 4.8 | 6.0 | 39.8 |
| **shot blasted** | 23.3 | 29.8 | 202 |
| **strong shot blasted** | 52.4 | 67.4 | 573 |

Tab. 2

| | | $\frac{\Delta k'_x}{k'_x(ref)}$ (%) | | |
|---|---|---|---|---|
| | | sanded | shot blasted | strong shot blasted |
| **mode** | **A1** | 2.0 | 2.1 | 4.5 |
| | **S1** | 2.0 | 2.5 | 4.6 |
| | **A2** | 2.1 | 2.2 | 4.4 |
| | **S2** | 2.1 | 2.4 | X |

Fig. 1

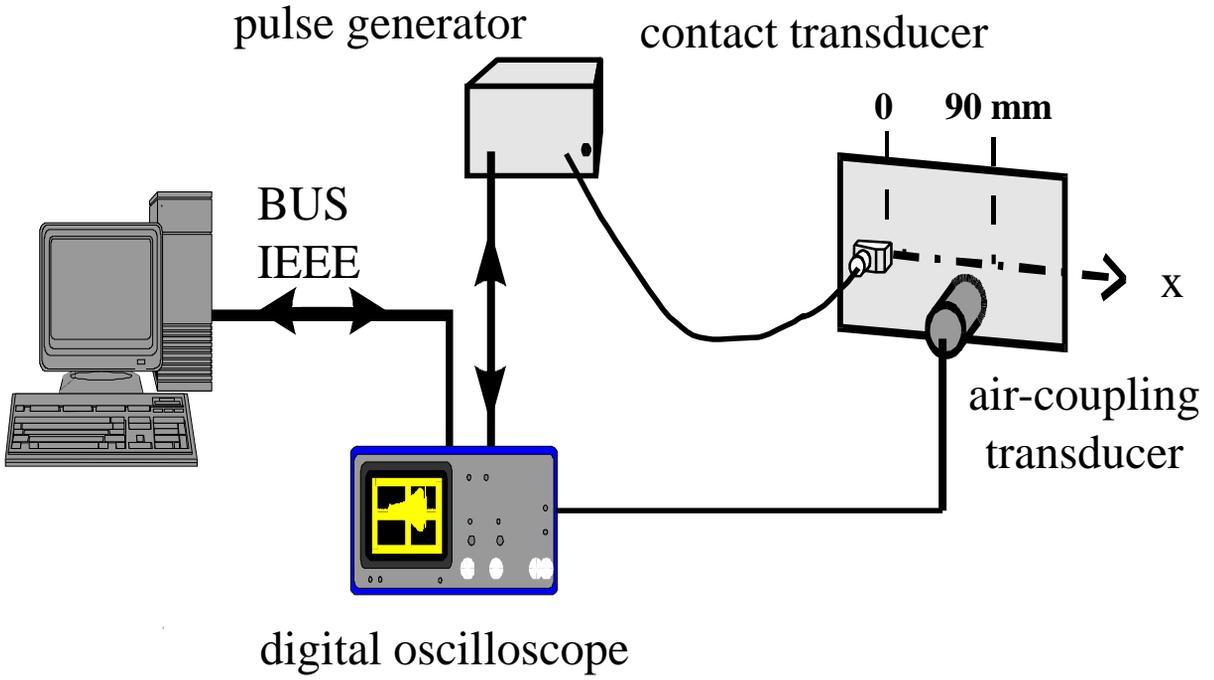

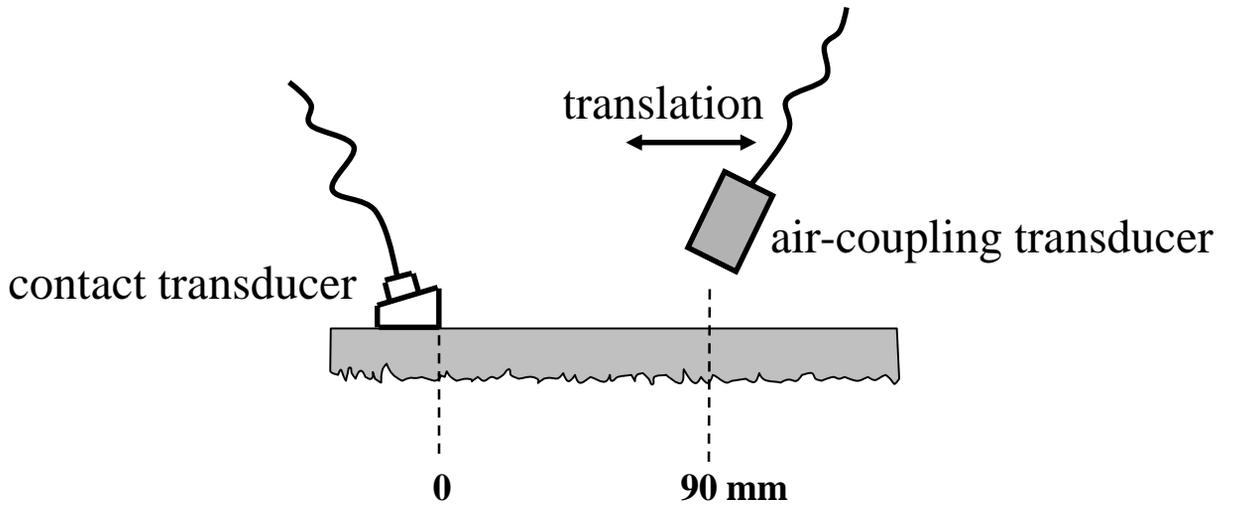

Fig. 2

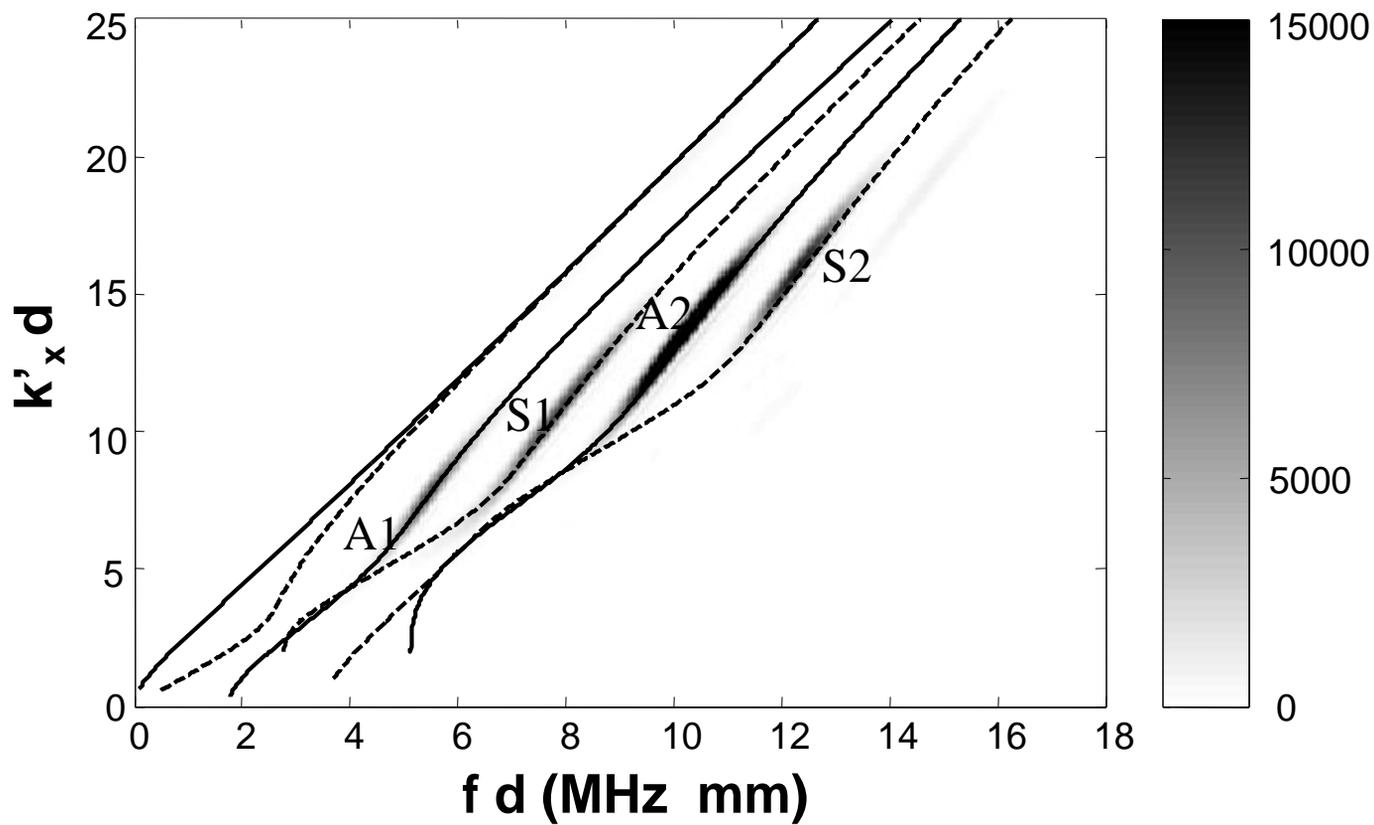

Fig. 3

a.

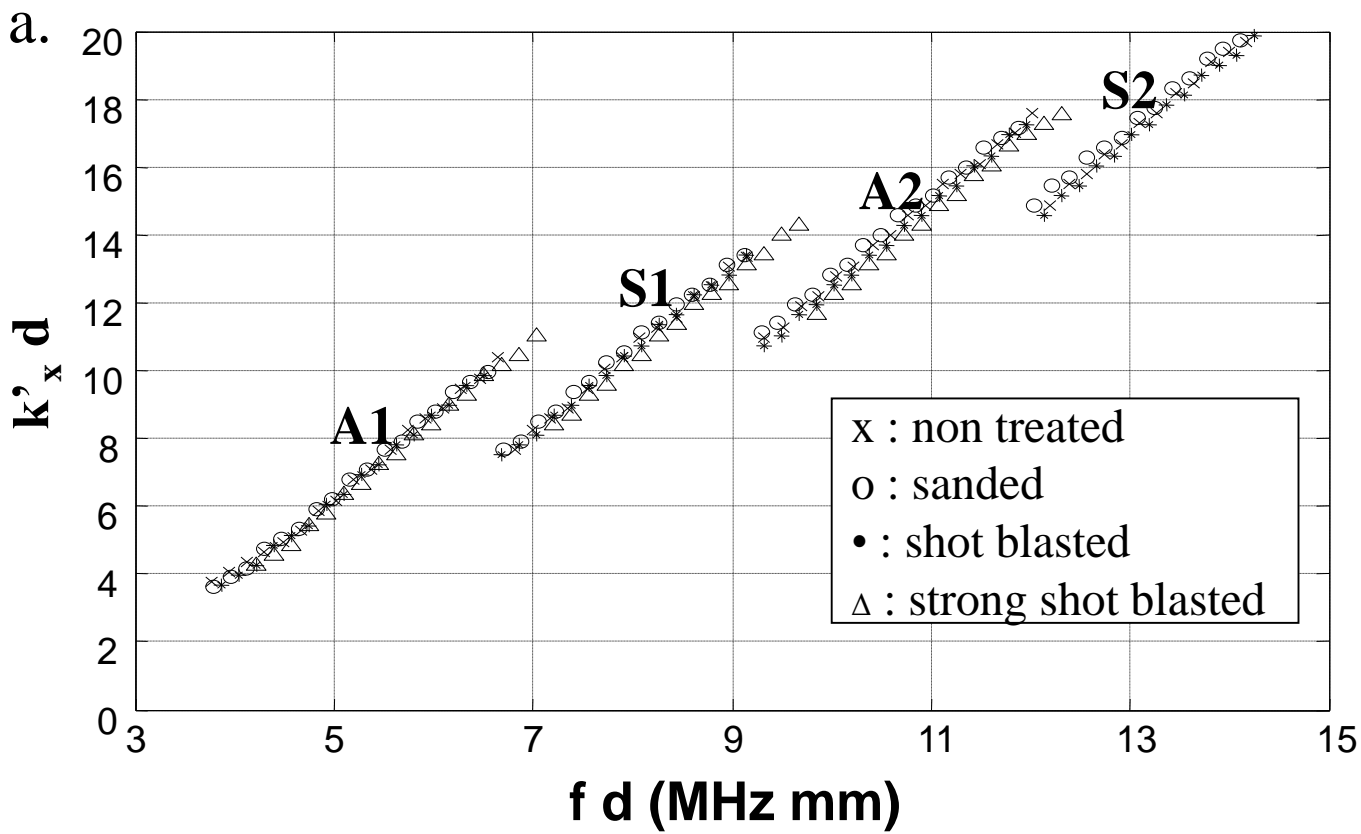

b.

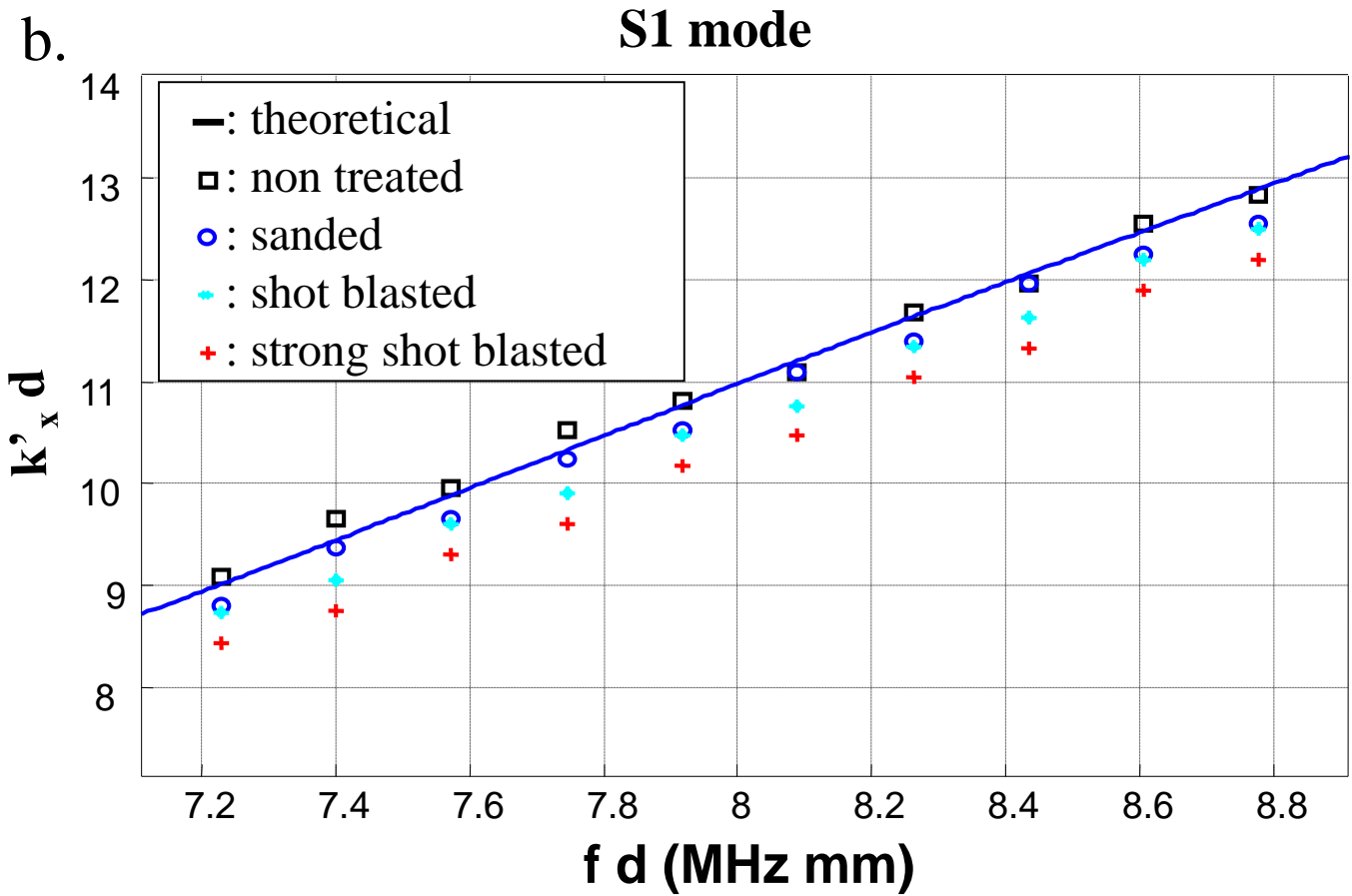

Fig. 4

a.
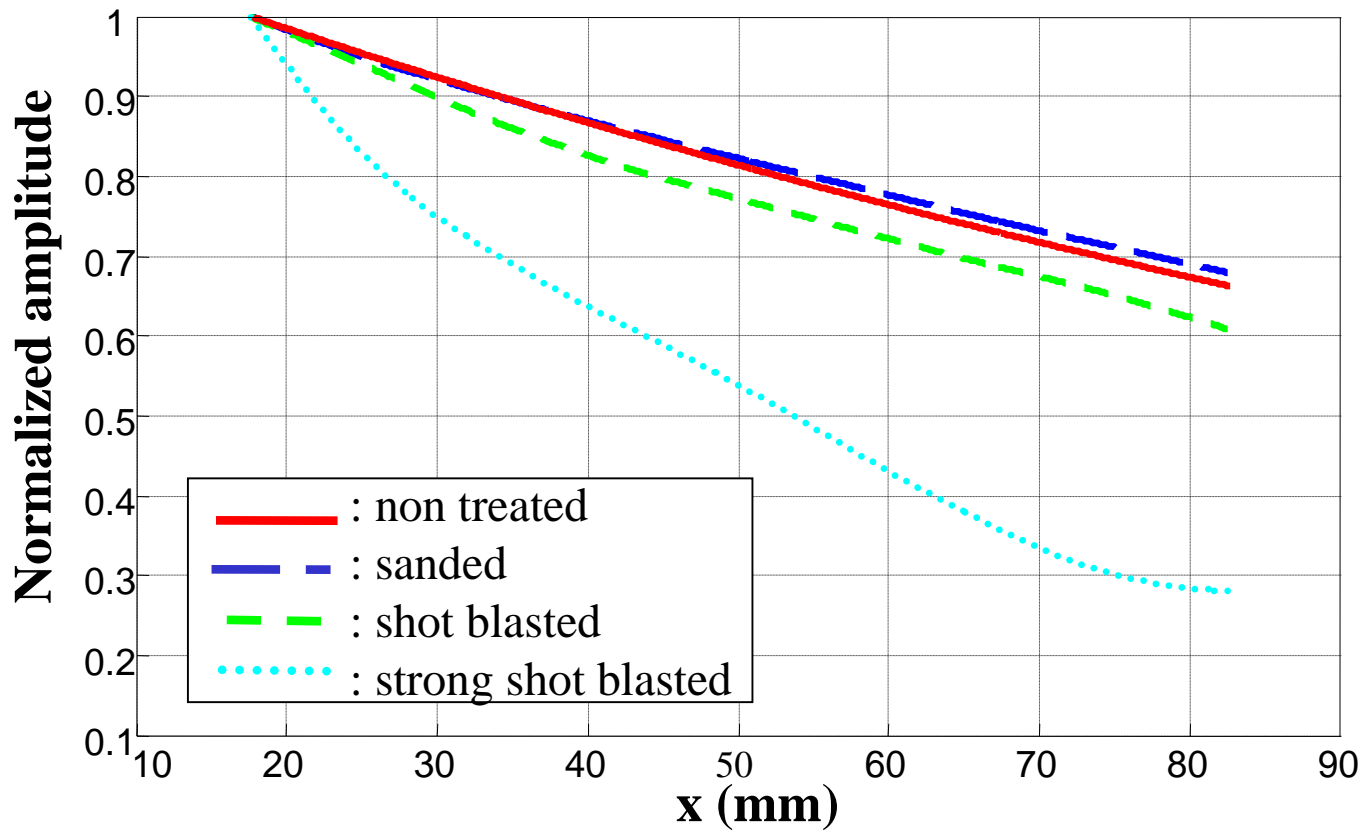

b.
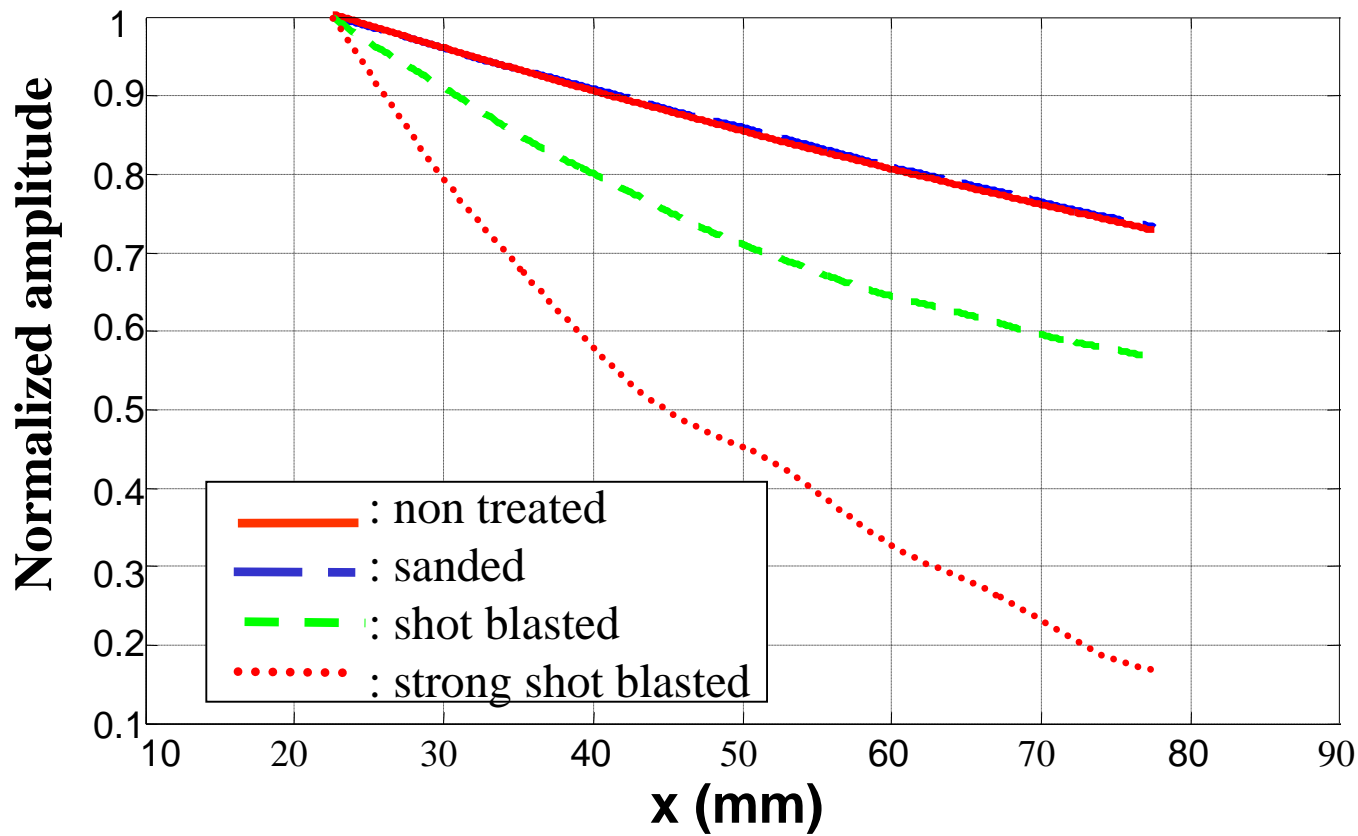

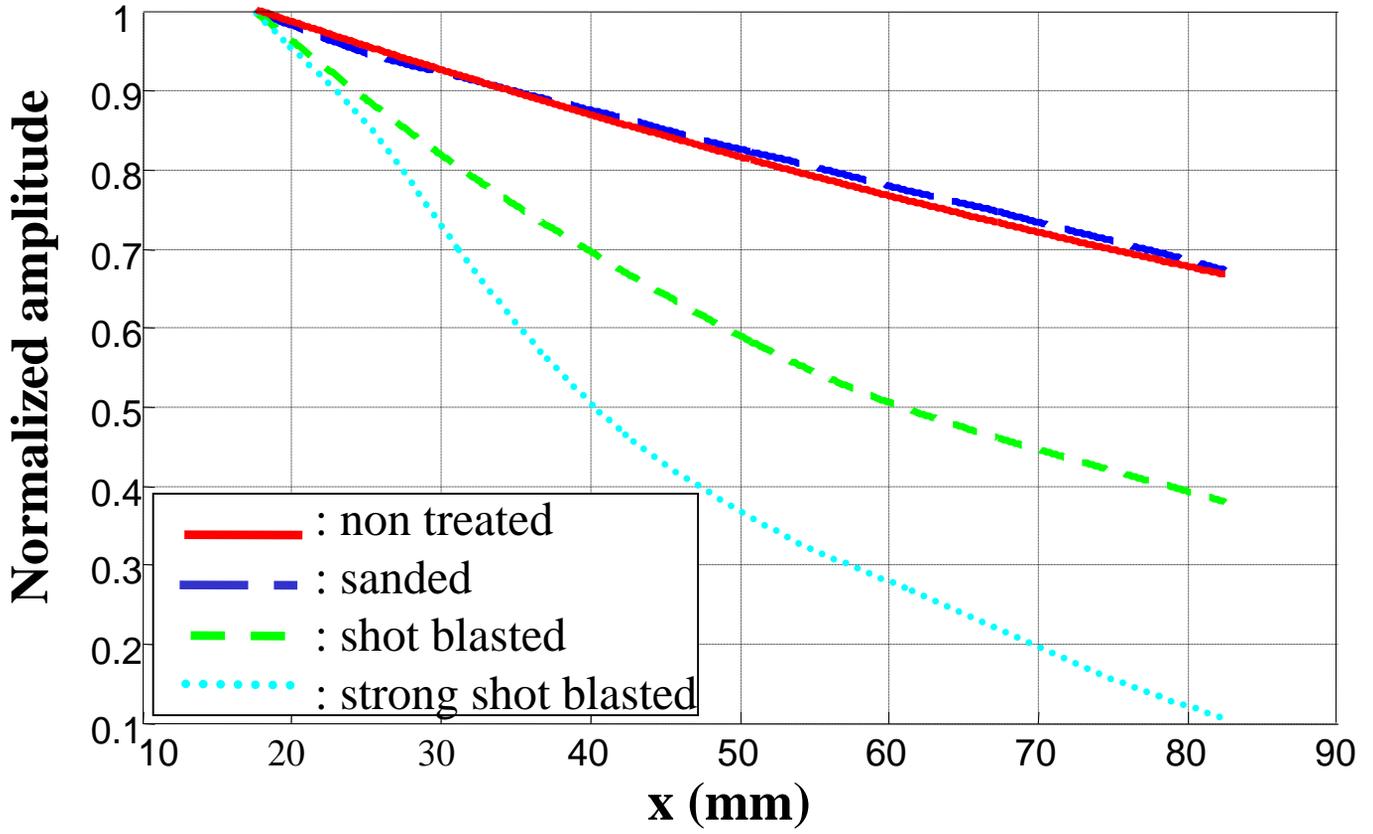

Fig. 5

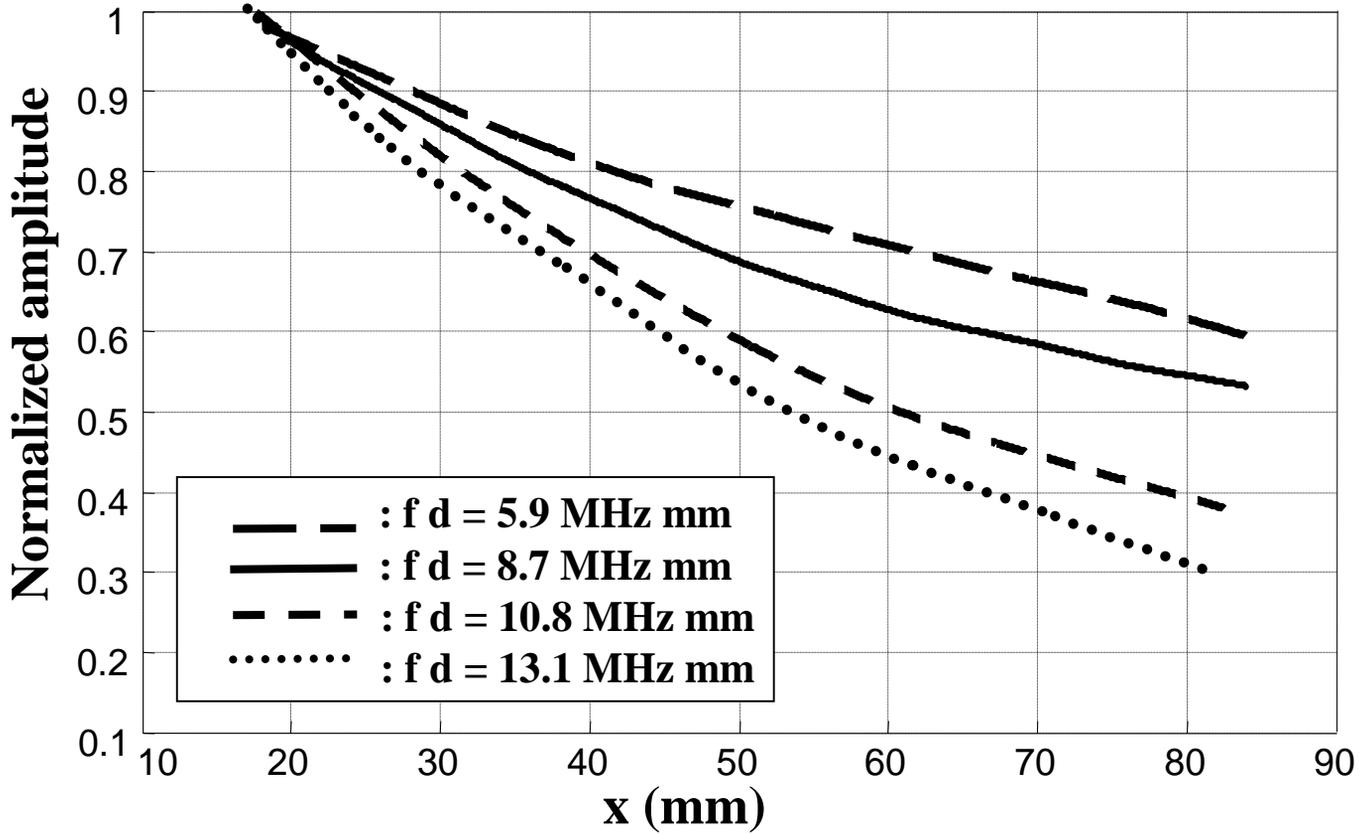

Fig. 6

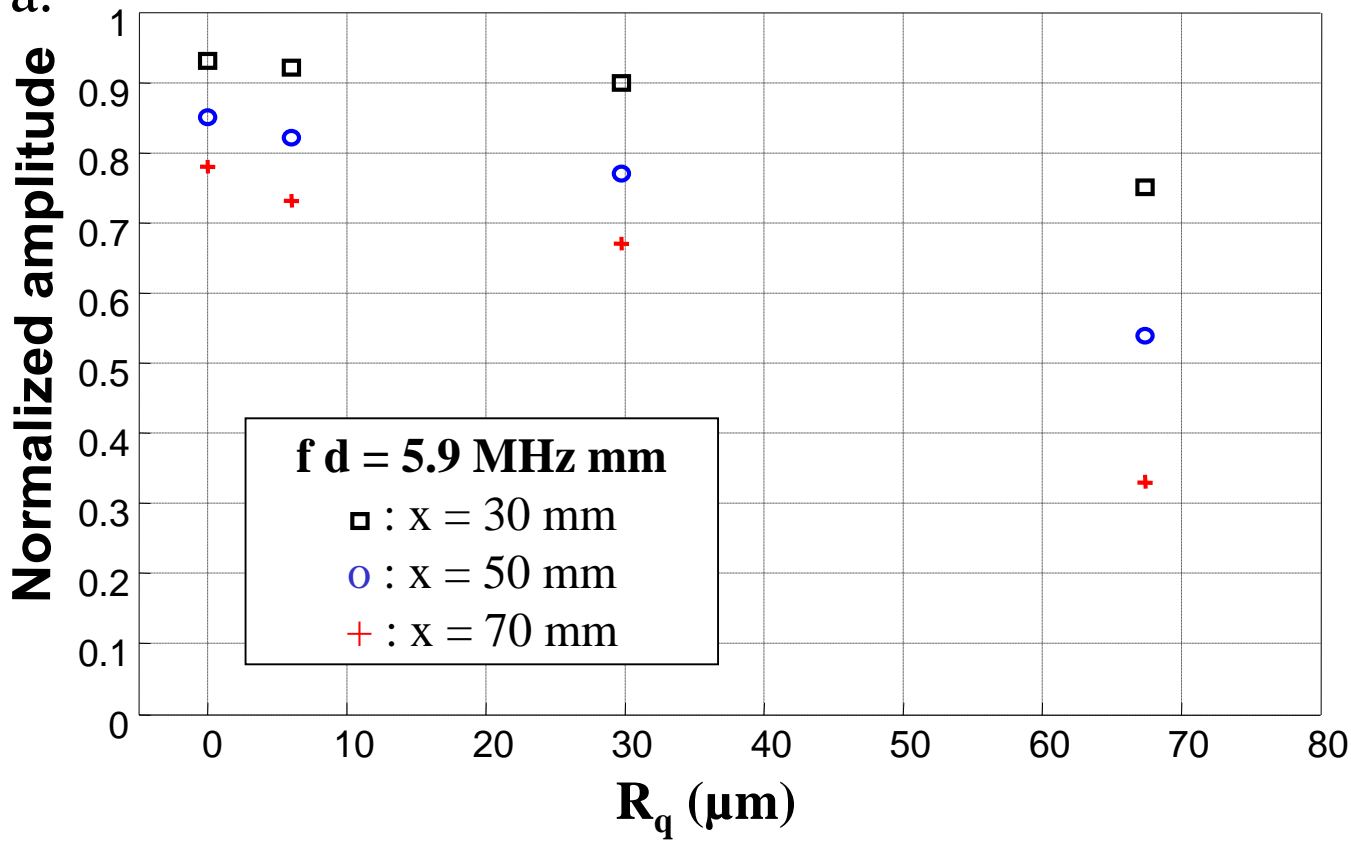

a.

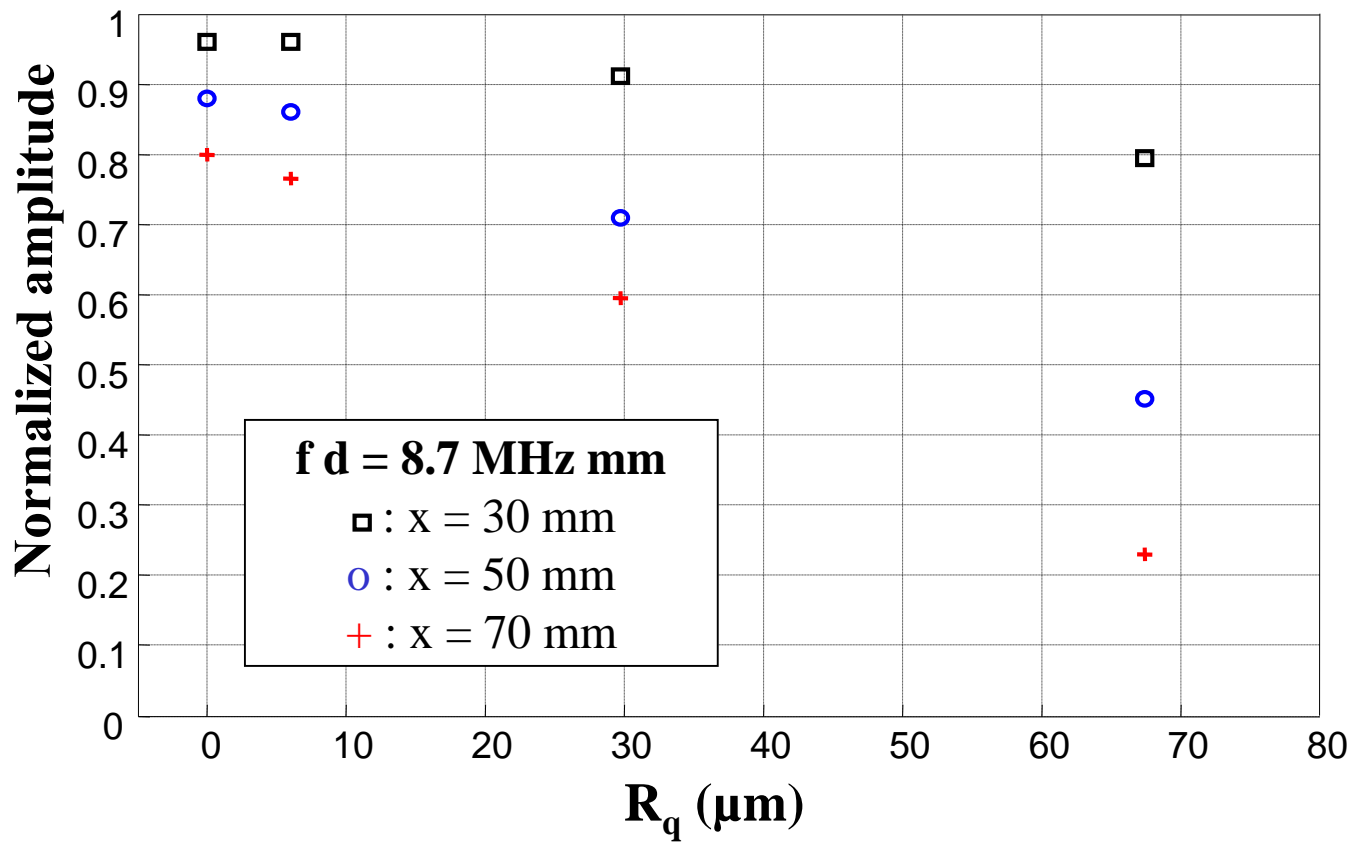

b.

c.

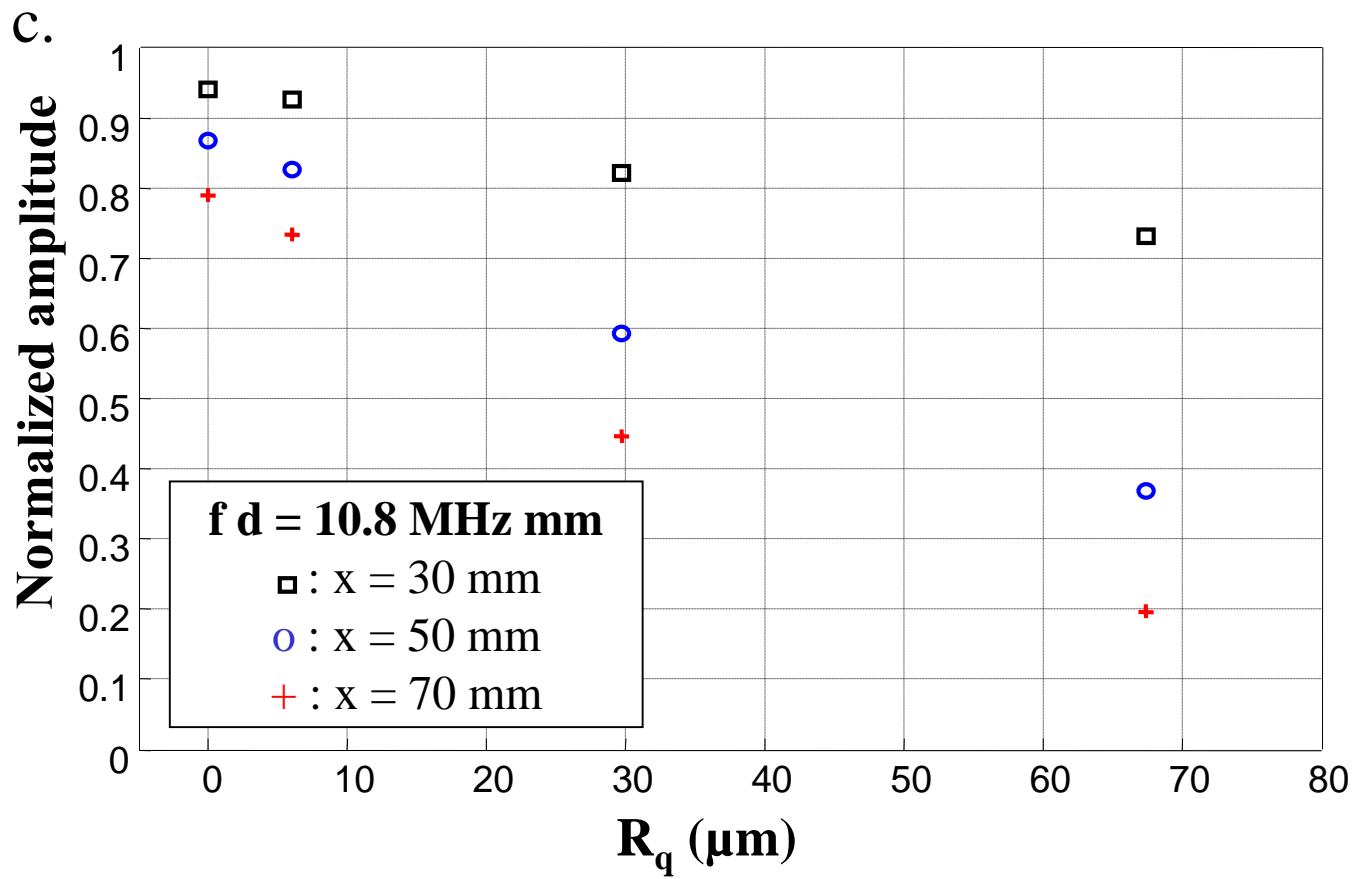